# On the Impact of Perceived Vulnerability in the Adoption of Information Systems Security Innovations

**Mumtaz Abdul Hameed**
Technovation Consulting and Training (Pvt) Ltd. 1/33, Chandhani Magu, Male'. 20003. Maldives.
E-mail: mumtazabdulhameed@gmail.com

**Nalin Asanka Gamagedara Arachchilage**
School of Engineering and Information Technology, University of New South Wales (UNSW Canberra), The Australian Defence Force Academy. Australia.
E-mail: nalin.asanka@adfa.edu.au

***Abstract***—A number of determinants predict the adoption of Information Systems (IS) security innovations. Amongst, perceived vulnerability of IS security threats has been examined in a number of past explorations. In this research, we examined the processes pursued in analysing the relationship between perceived vulnerability of IS security threats and the adoption of IS security innovations. The study uses Systematic Literature Review (SLR) method to evaluate the practice involved in examining perceived vulnerability on IS security innovation adoption. The SLR findings revealed the appropriateness of the existing empirical investigations of the relationship between perceived vulnerability of IS security threats on IS security innovation adoption. Furthermore, the SLR results confirmed that individuals who perceives vulnerable to an IS security threat are more likely to engage in the adoption an IS security innovation. In addition, the study validates the past studies on the relationship between perceived vulnerability and IS security innovation adoption.

## I. INTRODUCTION

Information System (IS) assets (information and computer resources) are at risk from a variety of threats, including virus, worms, Trojans, spyware, scare-ware, crime-ware, key-loggers, DDoS, pharming, phishing, etc. [4, 6]. Such attacks commonly referred to as 'IS security threats' are mainly intended to improperly disclose, modify or delete sensitive information and maliciously destruct and destroy computer resources [19].

Measures taken to thwart IS security threats include the installation of anti-phishing, anti-virus and anti-spyware software, setting up firewalls, maintaining and restricting access controls, using intrusion detection and prevention systems and by putting in encryption and content filtering software [28, 33]. These measures offer a technological or technical solution to the problem, but by no means reasonable to efficiently safeguard IS security threats completely [4, 26 47 61]. So as to endure with increased threats and to effectively protect IS assets, non-technical solutions such IS security policies have likewise been employed [46]. Research has established the view that organisations and individuals who opt for technical as well as non-technical measures to protect their IS assets are more likely to attain success in safeguarding IS resources [47]. Such technical and non-technical IS security measures may collectively referred as 'IS security innovations.'

Given that the security attacks are increasingly widespread and more organized than ever, it is important that the user be conscious about the vulnerabilities of the threats. IS security research use fear appeal to understand and administer IS security threats [26, 28]. A fear appeal is a persuasive message that attempts to arouse fear in order to divert behaviour through the threat of impending danger [49]. Fear appeal presents the potential risk, instructs the vulnerability to the risk, and then suggests the protective action. Founded on fear appeal concepts, Rogers [42] introduced a cognitive framework that predicts individual protective behaviour known as Protection Motivation Theory (PMT).

Although PMT was originally developed to explain the effects of fear appeals on health behaviours, PMT has often been used as the basis for many studies related to IS security [3, 26, 54, 56]. IS security literature exploited a number of other social cognitive theories including the Theory of Reasoned Action (TRA) by Fishbein and Ajzen [17] and the Theory of Planned Behaviour by Ajzen [1] to assess individual ability to prevent IS security threats. Among the various theories used in investigating IS security adoption behavioural research, PMT has been the most extensively used model. Anderson and Agarwal [3] suggest that PMT offers insights into the influence of various behaviours in IS security adoption. PMT's threat appraisal process evaluates the behaviour that prevents risks by cognitive traits that consists of perceived vulnerability and perceived severity of the threat. Perceived vulnerability is the individual's judgment regarding the likelihood that the threat would occur and perceived severity depicts the seriousness of the threat.

This research makes three main contributions to theory and practice. First, using a review of IS security literature, the research verifies the plausibility of the existing research on perceived vulnerability of IS threat on the adoption behaviour of IS security innovations. Secondly, the analysis carried out verifies the significance of examining the effect of perceived vulnerability on IS security adoption behaviour. Finally, the study approves the influence of perceived vulnerability of IS security threat for IS security adoption behaviour. We know to our cost this is the first attempt that attempts to review past literature that examine the relationship between perceived vulnerability of IS security threat on the adoption behaviour of IS security innovations.

The organisation of the remainder of this paper as follows. The 'Theoretical Background' section illustrates the basics of perceived vulnerability relating to IS security innovation adoption. In the subsequent section 'Research Questions', we presented four research questions for the study. The 'Research Methodology' section briefly discusses the method employed to examine the influence of the relationship between perceived vulnerability of IS security threat and IS security innovation adoption. In the following section, we presented the result obtained from the data analysis. The findings of the study were then clarified in the Discussion section. Finally, the conclusions were outlined with possible limitations of the study.

## II. RELATED WORKS

The focus of IS security is to protect and safeguard organization's IS assets from vulnerabilities [2]. The main challenge for organization's IS security is to protect unauthorized access of information sources [16] and to defend computer resources against malicious attacks. In the context of IS security, individuals within that organisation are likely to feel some effects if an organisation experiences IS security threat. Furthermore, it is practical to assume that an individual who perceives that their IS assets are vulnerable is more likely adopt IS security measures as a 'protection behaviour.'

According to Lee et al. [32] the term 'protection behaviour' is an adaptation from PMT. Drawing from the expectancy-value theories and the cognitive processing theories, Rogers [43] developed PMT to help explain a fear appeal. The principal view of the PMT is that individuals tend to develop an intention to protect against an attack if he or she is susceptible to any risk. Also, the PMT explains how people cope with potential threats [42]. The theory has been widely used to explain and predict a variety of protective behaviours, including IS security threats. In a meta-analysis study, Floyd et al. [18] described PMT as one of the most powerful explanatory theories predicting individual intentions to safeguard IS security measures.

Several extant researches has used PMT and found the assumptions very useful in predicting individuals IS security behaviours for both individual and organizational contexts [3, 33, 41]. PMT outlines the action of dealing with a potential threat as the result of two appraisal processes, namely a threat appraisal and a coping appraisal. Threat appraisal of PMT suggests that individuals will consider both perceived vulnerability and perceived severity of a threat, when faced with a risk. In the context of threat assessment, perceived vulnerability is the degree to which an individual believes a threat will occur to him or her [32].

Perceived vulnerability concerns the susceptibility a person has to a threat. Individuals with higher degree of vulnerability are more conscious about IS security protection. Lee and Larsen [33] suggest that perceived vulnerability has significant impact on individual intention to adopt security tools. In general, perceived vulnerability of IS security threats have found to influence IS security adoption, behaviour [28, 41, 46, 57]. PMT also predicts that perceived vulnerability positively influences IS security threat avoidance intention. A number of studies suggest that an individual who perceived that his or her IS assets are vulnerable to an IS attack, they are more likely to take protective action [24, 38, 53]. Over and above, the findings of the study by Ng et al. [41] determined that perceived vulnerability of IS threat is as an important antecedence for user computer security behaviour. Similarly, Chenoweth et al. [11] found perceived vulnerability of IS security threat has a positive effect on behavioural intention to adopt anti-spyware software. A study that examines the adoption of IS privacy protection measures in social networking sites, Mohamed and Ahmad [39] demonstrated that the perceived vulnerability of an IS threat enabled the adoption of privacy measures to protect individuals' information privacy. While, the majority of studies found perceived vulnerability of IS threat to have a positive influence [11, 35], some studies found a negative effect [31, 40]. Nevertheless, it is hypothesized that individuals who perceive themselves as vulnerable to IS security breaches will be more motivated to protect their IS assets.

Several researches have examined the relationship between perceived vulnerability and IS security innovation adoption [54, 56]. However, the relationship between perceived vulnerability and IS security innovation adoption has yielded inconsistent results. It is important to address the reasons why there has been so much inconsistency in past studies in establishing the relationship. In order to corroborate the relationship and to clarify the mixed and inconsistent findings, this research attempts to investigate the practices used in the research that examine perceived vulnerability of IS security threat on the adoption behaviour of IS security innovations. To this end, the study reviewed past literature that examine the relationship between perceived vulnerability and IS security adoption behaviour. This allows the validation of the past research processes and hence, clarify the inconsistencies that might exist among the studies.

## III. RESEARCH QUESTIONS

This paper considered the existing IS security literature to analyse the proceedings of the research that examine the influence of perceived vulnerability of an IS security threat on the adoption behaviour of IS security innovations. In particular, the analysis focused on investigating, the following research questions:

RQ1: What are the demographics of the extant studies of perceived vulnerability of IS security threat on the adoption, behaviour of IS protective measures, including the year of study, sample groups, sample size, countries?

RQ2: What are the main theories used in the existing studies of perceived vulnerability on the adoption behaviour of IS security measures?

RQ3: Is there a difference in investigating the perceived vulnerability for different types of IS security innovations?

RQ4: What are the results of the studies that examine the relationship between perceived vulnerability of threat and the adoption of IS protective measures?

## IV. RESEARCH METHODOLOGY

Finding of an individual study is not sufficient to generalise on a particular issue, while combining the findings of a number of independent studies on a subject, results in a better overall outcome [20]. One technique used to identify, analyse and interpret all available evidence related to a particular research question is a Systematic Literature Review (SLR). To meet our research objectives and to address the research questions, we carried out a SLR to analyse the past examinations of perceived vulnerability of IS security threat for the adoption behaviour of IS security innovations. The use of SLR procedure enabled the study to aggregate the findings of past studies on perceived vulnerability of IS security threats that influence the adoption behaviour of IS security protective measures to obtain an overall conclusion regarding the relationships. In addition, we used SLR to ensure that the research accurately and thoroughly covered the findings of past studies. Also, compare to other literature review approaches, SLR is a better choice in terms of transparency, as other researchers can straightforwardly replicate the findings of the study.

To ensure a thorough coverage of academic articles related to IS security behaviour, we conducted an extensive literature search of IS Journals using Google Scholar and digital databases including Web of Science, IEEE Xplore, Science Direct (Elsevier), ACM Digital Library, Wiley Online Library, ProQuest, EBSCO, Springer LINK and Emerald Management Xtra. These sources contain ample high-quality journal articles and conference papers.

To determine which of the articles were really relevant to the research objectives the study established, an inclusion and exclusion conditions. The study selection criteria for the SLR were: (C1) it should be an empirical study on IS security behaviour; (C2) the study should examine perceived vulnerability of IS security threat as one of the independent variables, and finally (C3) the study should examine the relationship between perceived vulnerability of IS threat and adoption behaviour of IS security innovations.

The initial search yielded 493 citations by following inclusion and exclusion criterion C1. The abstracts of all 493 were manually scanned to identify if the articles examine perceived vulnerability to accomplish the inclusion and exclusion criterion C2. A total of 139 articles were found potentially relevant. These articles were further subjected to inclusion and exclusion criterion C3 and a total of 38 articles were found eligible for the SLR after meeting all three criteria. These 38 studies examine the effect of perceived vulnerability of IS security threat for the adoption behaviour of IS security innovations. The studies used different terminology to depict perceived vulnerability. Considering the definition of the term, we considered perceived susceptibility, the likelihood of a threat, the certainty of a security breach, risk perception and probability of a security breach as perceived vulnerability.

## V. RESULTS

We conducted a statistical analysis using frequencies and percentages to combine and summarize the variables collected.

### A. Disribution of Studies by Year

Table 1 shows the literature distribution by publication year of the studies. Data from the SLR shows that perceived vulnerability of IS security threat has been considered in the IS security adoption literature since 2008. The academic discussion of perceived vulnerability on IS security adoption has mostly taken place during the past 10 years.

The distribution of studies by publication year suggests that examining the perceived vulnerability of IS security threat for the adoption, behaviour of IS security innovation is an increasingly emerging discourse. Also, SLR confirms that the perceived vulnerability of IS security threat for the adoption, behaviour of IS security innovation is still an active IS tract, as there were 9 articles published in 2017 and 7 articles so far in 2018. The result of this analysis provided part of the clarification for RQ1.

Table 1. Literature Distribution by Publication Year

| Year | No. of Studies |
|---|---|
| 2008 | 2 |
| 2009 | 3 |
| 2010 | 4 |
| 2011 | 0 |
| 2012 | 5 |
| 2013 | 1 |
| 2014 | 3 |
| 2015 | 0 |
| 2016 | 4 |
| 2017 | 9 |
| 2018 | 7 |

*B. Distribution of Sample Groups in the Studies*

In response to RQ1, we analysed the different sample groups which the studies employed. Table 2 illustrates the different sample subjects used in examining the perceived vulnerability of IS security threat for the adoption of IS security innovations. Results suggest that the majority of studies conducted their studies by engaging individuals, adopting convenience sampling.

Table 2. Distribution of Sample Groups used in the Studies

| Subject Groups | No of Studies |
|---|---|
| Individual | 22 |
| Organisation | 2 |
| Student | 11 |
| Mixed | 3 |
| None | 0 |

Table 3. Distribution of Sample Size of the Studies

| Description | No. of. Studies |
|---|---|
| Studies with sample | 38 |
| Smallest sample size | 77 |
| Largest sample size | 1201 |
| Sample Size 0 - 100 | 1 |
| Sample Size 101 - 200 | 8 |
| Sample Size 201 - 300 | 8 |
| Sample Size 301 - 400 | 9 |
| Sample Size 401 - 500 | 3 |
| Sample Size 501 - 600 | 4 |
| Sample Size 601 - 700 | 3 |
| Sample Size 701 - 800 | 0 |
| Sample Size 801 - 900 | 0 |
| Sample Size 901 - 1000 | 1 |
| Sample Size > 1000 | 1 |

*C. Distribution of Sample Size in the Studies*

The SLR also analysed sample size of the reviewed studies to address RQ1. All 38 studies considered in the SLR utilised survey methodology using different sample groups. Overall 38 studies used 13,768 participants, with an average sample size of 362. As depicted in the Table 3, the study that used the smallest sample is 77 and the largest study employed 1201 participants.

*D. Distribution by Country*

As the final appraisal for RQ1, we analysed the distribution of different culture in the reviewed literature. Each of the reviewed articles considered in the SLR conducted their investigation in a single locality, which allows to account country settings. Table 4 visually indicates that almost 60% of the studies conducted in the USA and the literature represent Asia, Africa, Europe and North America.

Table 4. Distribution of Country of the Studies

| Country | No. of Study |
|---|---|
| Australia | 1 |
| Canada | 2 |
| China | 1 |
| Finland | 2 |
| Hong Kong | 1 |
| Indonesia | 1 |
| Malaysia | 1 |
| Netherlands | 1 |
| Singapore | 1 |
| South Africa | 1 |
| South Korea | 1 |
| Taiwan | 2 |
| USA | 23 |

*E. Theories used in the Reviewed Studies*

In response to RQ2, we analysed the theoretical foundation for each reviewed literature. Table 5 shows the different theoretical model exploited in the studies considered in the SLR. PMT is the most widely used theory to determine the relationship between perceived vulnerability of IS security threat and adoption behaviour of IS security innovations. More than three-fourth of the reviewed studies used PMT or PMT integrated with other theories. Reviewed literature suggests that the TPB and Technology Threat Avoidance Theory (TTAT) are among the other theories utilised in examining the perceived vulnerability for the adoption behaviour of IS security innovations.

Table 5. Different Theories used in the Studies

| Theories | No. of Studies |
|---|---|
| Protection Motivation Theory (PMT) | 29 |
| Theory of Planned Behaviour (TPB) | 4 |
| Technology Threat Avoidance Theory (TTAT) | 3 |
| Theory of Reasoned Action (TRA) | 2 |
| Decomposed Theory of Planned Behaviour (DTPB) | 2 |
| Deterrence Theory (DT) | 2 |
| Technology Acceptance Model (TAM) | 2 |
| Social Cognitive Theory (SCT) | 1 |
| Health Belief Model (HBM) | 1 |
| Rational Choice Theory (RCT) | 1 |

*F. Types of Protective Measures*

According to the classification of Zmud [60] we defined the type of innovations as a process and product IS security measures. Accordingly, process IS security innovations involve establishing a new system, method or policies that changes the IS security operational processes, whereas product IS security innovations are new products introduced to enhance IS security. Different factors determine the adoption of process and product innovations and the extent to which these factors impacts on the adoption process [51].

Table 6. Distribution of Studies using Different Security Innovations

| Description | Process | Product |
|---|---|---|
| No of Studies | 20 | 18 |
| Total sample size | 0 | 0 |
| Sample Group | | |
| Individual | 13 | 9 |
| Organisation | 2 | 0 |
| Student | 5 | 6 |
| Mixed | 0 | 3 |
| None | 0 | 0 |
| Theories used | | |
| Protection Motivation Theory (PMT) | 15 | 14 |
| Theory of Planned Behaviour (TPB) | 4 | 0 |
| Technology Threat Avoidance Theory (TTAT) | 1 | 2 |
| Theory of Reasoned Action (TRA) | 1 | 0 |
| Decomposed Theory of Planned Behaviour (DTPB) | 2 | 0 |
| Deterrence Theory (DT) | 2 | 2 |
| Technology Acceptance Model (TAM) | 1 | 1 |

We differentiate the reviewed studies into process and product IS security innovations and examine some demographics including sample size, sample groups for each group of these studies. Also, we examine if there is any difference in the application of theories for the studies that examine the process and product IS security innovations. Table 6 highlights the variance of studies using both process and product IS security innovations. The result of this analysis would provide clarification for RQ3. The results showed that the average sample size used for process IS security innovation studies (368 participants) and the product IS security innovation studies (355 participants) are evenly matched. Also, it is evident from the results that most of process IS security innovation studies utilise non-students as subjects, whereas, most of product IS security innovation studies employed both students and non-students. Both the process and product IS security innovation studies predominantly capitalised PMT as the theoretical basis in the investigations.

*G. Significnce*

The relationship between independent and dependent variables is usually evaluated in term of 'test of significance', highlighting their relationship. Aggregation of 'Test of significance' and various other 'effect sizes' such as correlation co-efficient provided by quantitative studies provides an overall outcome of a relationship [23]. Effect size when considered in terms of significance is frequently referred as weak, moderate or strong significance [20]. Hunter et al. [27] and Hameed and Counsell [21], however, suggested that aggregation of 'test of significance' results from different studies could produce a misleading outcome. This is because there is no rule for determining the value of the correlation coefficient that interprets as weak, moderate or strong significance.

Table 7. Aggregated test of Significance for the Studies

| Significance | No. of Studies |
|---|---|
| Insignificant (0.00 to ±0.09) | 4 |
| Weak Significance (0.10 to ±0.29) | 13 |
| Moderate Significance (0.30 to ±0.49) | 6 |
| Strong Significance (0.50 to ±0.69) | 4 |
| Very Strong Significance (0.70 to ±0.89) | 3 |
| Perfect (0.10 to ±1.00) | 0 |

For this study, we extracted correlation co-efficient values of the relationship between the perceived vulnerability of IS security threats and the adoption behaviour of IS security innovations. By interpreting each of the correlation values under a single classification, we obtained the test of significance for the assessment. We adopted the correlation value referred by Hameed and Counsell [20] and Hameed and Counsell [22], i.e.: a correlation value between 0 and ±0.09 as insignificant, ±0.10 and ±0.29 as weak significance, ±0.30 and ±0.49 as moderate significance, ± 0.5 and ± 0.69 as strong significance, ±0.70 and ±0.89 as the very strong significance and ±0.9 and ±1.0 near perfect.

Among the 38 studies considered in the SLR, 30 studies provided correlation co-efficient for the relationship between the perceived vulnerability of IS security threats and the adoption behaviour of IS security innovations. According to the above classification, we coded the correlation co-efficient of individual studies and aggregated the resulting test of significance to obtain the overall assessment of the relationship. Table 7 summarizes the results of an aggregated test of significance and illustrates answer the RQ4.

## VI. Discussion

This SLR aimed to evaluate the practice involved in examining the influence of individual perceived vulnerability of IS security threat on the adoption of IS security innovation. The results highlighted the significance of examining the perceived vulnerability of IS security threat in the adoption behaviour of IS security innovation.

The SLR results of the distribution of studies by publication year suggest that researchers have started examining the effect of the perceived vulnerability of IS security threat on the adoption of IS security innovations since 2008. With the launching of Facebook in 2004, online social media and social networking became a mainstream concept for any type of interaction with other people or businesses, when communication is important. As such, individuals and organisations put their private and confidential information online, offering a huge opportunity for cyber criminals to exploit. Social media emerge as a target for scams; putting individual and organisational data at risk. Correspondingly, IS security adoption studies has become increasingly attracted among the scholarly researchers.

The SLR findings showed that the research on the relationship between the perceived vulnerability of IS security threats and adoption of IS security innovations used convenience sampling and have employed students and non-students. Use of student subjects for experimental research has been widely criticised for (1) having little external validity and generalisability; (2) whether forced to serve as participants, and (3) biased in age, experience, and intellectual ability [36]. However, the studies reviewed in the SLR provided no justification for their chosen subject sample nor did acknowledge any limitations for the use of students' sample. Hence, studies that examined the effect of perceived vulnerability of IS security threats on adoption of IS security innovations take no distinction for student and non-student sample.

The results of SLR showed that the average sample size of the studies are 362 participants. So, with a margin of error of 5% and confidence level of 95%, the average population size of the studies is approximately 6300. For an infinite population size at 5% margin of error and 95% confidence level, the sample size required is 385. On that account, the sample size used in the majority of the reviewed studies for the SLR deemed appropriate. It manifests the soundness of the selected studies for the SLR and the correctness of the results of the reviewed studies that examine the relationship between perceived vulnerability of IS security threats and adoption behaviour of IT security innovations. A study that has a sample size which is too small may have an unrealistic chance of yielding a useful information, while larger sample sizes have the obvious advantage more accurate mean values and a smaller margin of error.

In order to identify if culture moderates, the relationship between perceive vulnerability of IS security threats and adoption of IS security innovations, we analysed the variance of the locality of the study settings. Deans et al. [15] states that culture influences usage of IS in different countries. The SLR represents a diverse culture, hence, the existing literature on the influence of perceived vulnerability of IS security threats for the adoption behaviour of IS security innovations reasonably represents blends of discrete cultures.

The SLR explored the theoretical foundation exploited in examining the perceived vulnerability of IS security threats on the adoption behaviour of IS security innovations by the reviewed studies. The result of the SLR identified PMT as the principal model. PMT has widely been exploited to analyse the behaviours to avert the IS security attacks. Apart from PMT, SLR identified TPB and TTAT as other models utilised in examining the effect of the perceived vulnerability of IS security threats on the adoption behaviour of IS security innovations.

The SLR showed that there was no difference in the investigation of perceived vulnerability of IS security threats for either process or product IS security innovations. IS literature suggests that the adoption behaviour of product and process innovation vary significantly [20] and that firms often adopt mixed modes of innovation, meaning that they combine product and process innovations.

Finally, the SLR analysed the test of significance for the relationship between perceived vulnerability of IS security threats on the adoption, behaviour of IS security innovations. In terms of test of significance, just about 96% of the studies found either weak, moderate or strong significance perceived vulnerability of IS security threats and the adoption, behaviour of IS security innovations. Hence, consistent with PMT, the results confirm that perceived vulnerability is a significant predictor of IS security adoption behaviour.

## VII. Conclusions

A good number of studies examines the influence of perceived vulnerability of IS security threat for the adoption behaviour of IS security innovations. This study aggregated 38 extant literatures that examined perceived vulnerability of IS security threat for the adoption behaviour of IS security innovations. The study conducted a SLR methodology to identify, analyse and interpret all the existing findings on the relationship between perceived vulnerability and IS security innovation adoption behaviour. We know to our cost this is the first attempt a SLR procedure has been carried out to report on the investigation of perceived vulnerability of IS security threat for the adoption behaviour of IS security innovations.

The SLR found that perceived vulnerability of IS security threat has consistently been exploited in the IS security innovation adoption literature during the past 10 years. SLR also revealed that studies examining perceived vulnerability of IS security threat for IS security innovation adoption is still increasing. The review also found that the dominant theory that underpin the relationship between perceived vulnerability of IS security threat and adoption behaviour of IS security innovations is PMT. Finally, the aggregated results of the past studies confirmed that perceived vulnerability of IS security threat is a significant determinant of IS security innovations adoption.

Use of SLR principles in this study was found highly valuable. It increased the overall knowledge regarding perceived vulnerability of IS security threat for IS security innovations adoption behaviour. In addition, the use of systematic literature review enabled to highlight methodological variations that exists when examining perceived vulnerability of IS security threat for IS security innovations adoption behaviour.

The principle implications of this research seem to be that examining perceived vulnerability of IS security threat for IS security innovations adoption behaviour is still an active research theme. Studies need explore perceived vulnerability of IS security threat for different research setting and for varied IS security innovations. The most important theoretical implication of this study is that individuals need to consider the vulnerability of IS security threat more passionately. We suggest that organizations create appropriate education, training and security awareness programs that ensure employees possess up-to-date knowledge of IS security as well as facilitate conditions that will improve their view regarding the vulnerability of IS threats.

This study has certain limitations and in interpreting the results of this study, its limitations need considering. The major limitation of this analysis was that the searching websites undermined the objectivity of the search, and relevant websites may have excluded. This means that a number of pertinent studies could have missed. Also, the search terms used for finding the relevant studies were only based on variable terminology that has been used in the literature. Therefore, our findings with regards to terminology used to describe perceived vulnerability and methods used analyse the relationship may not be comprehensive or wholly representative. Another limitation is the inadequacy of studies that examined the perceived vulnerability of IS security threats and the adoption behaviour of IS security innovations. The result could have been more accurate and better explained, if the SLR conducted with more studies. Despite these limitations, we feel that this work reflects the variation in methodology used in examining the perceived vulnerability of IS security threats and the adoption behaviour of IS security innovations. Furthermore, the study clearly underlines the demographics, methodology and reporting of the research that examine perceived vulnerability of IS security threats and the adoption behaviour of IS security innovations.

REFERENCES

[1] I. Ajzen, "The Theory of Planned Behaviour," *Organizational Behaviour and Human Decision Processes*, vol. 50, pp. 179–211, 1991.
[2] A. Alshboul, "Information Systems Security Measures and Countermeasures: Protecting Organisational Assets from Malicious Attacks," *Communications of the IBIMA*, pp. 9p, 2010.
[3] C. L. Anderson and R. Agarwal, "Practicing Safe Computing: A Multimedia Empirical Examination of Home Computer User Security Behavioral Intentions," *MIS Quarterly*, vol. 34, no. 3, pp. 613-643, 2010.
[4] N. A. G. Arachchilage and M. A. Hameed, "Integrating Self-efficacy into a Gamified Approach to Thwart Phishing Attacks," In: *The Proceedings of 5th International Conference on Cybercrime and Computer Forensics (ICCCF)*, 2017.
[5] S. Aurigemma and T. Mattson, T. "Do it OR ELSE! Exploring the Effectiveness of Deterrence on Employee Compliance with Information Security Policies," In: *The Proceedings of 20th American Conference of Information Systems (AMCIS)*, 2014.
[6] M. N. Banu and S. M. Banu, "A Comprehensive Study of Phishing Attacks," *International Journal of Computer Science and Information Technologies*, vol. 4, no. 6, pp. 783-786, 2013.
[7] F. Béla nger, S. Collignon, K. Enget and E. Negangard, "Determinants of Early Conformance with Information Security Policies," *Information and Management*, vol. 54, pp. 887-901, 2017.
[8] B. Bulgurcu, H. Cavusoglu and I. Benbasat, "Information Security Policy Compliance: An Empirical Study of Rationality-Based Beliefs and Information Security Awareness," *MIS quarterly*, vol. 34, no. 3, pp. 523-555, 2010.
[9] A. J. Burns, C. Posey, T. L. Roberts and P. B. Lowry, "Examining the Relationship of Organizational Insiders' Psychological Capital with Information Security Threat and Coping Appraisals," *Computers in Human Behavior*, vol. 68, pp. 190-209, 2017.
[10] Y. Chen, "Examining Internet Users' Adaptive and Maladaptive Security Behaviors Using the Extended Parallel Process Model," In: *The Proceedings of International Conference of Information Systems (ICIS)*, 2017.
[11] T. Chenoweth, R. Minch and T. Gattiker, "Application of Protection Motivation Theory to Adoption of Protective Technologies," In: *The Proceedings of the 42nd Hawaii International Conference on System Sciences (HICSS)*, 2009.
[12] V. Cho and W. H.Ip, "A Study of BYOD Adoption from the Lens of Threat and Coping Appraisal of its Security Policy," *Enterprise Information Systems*, vol. 12, no. 6, pp. 659-673, 2018.
[13] H. Chou and C. Chien, "An Analysis of Multiple Factors Relating to Teachers Problematic Information Security

Behavior," *Computers in Human Behavior*, vol. 65, pp. 334-345, 2016.

[14] R. E. Crossler, "Protection Motivation Theory: Understanding Determinants to Backing up Personal Data," In: *The Proceedings of the 43rd Hawaii International Conference on System Sciences (HICSS)*, 2010.

[15] C. P. Deans, K. R. Karawan, M. D. Goslar, D. A. Ricks and B. Toyne, "Identification of Key International Information Systems Issues," *Journal of High Technology Management Review*, vol. 2, no. 1, pp. 57-81, 1991.

[16] Y. S. Feruza and T. Kim, "IT Security Review: Privacy, Protection, Access Control, Assurance and System Security," *International Journal of Multimedia and Ubiquitous Engineering*, vol. 2, no. 2, pp. 17-32, 2007.

[17] M. Fishbein and I. Ajzen, "Belief, Attitude, Intention and Behaviour: An Introduction to Theory and Research," Addison-Wesley, Reading, MA, 1975.

[18] D. L. Floyd, S. Prentice-Dunn and R. W. Rogers, "A Meta-Analysis of Research on Protection Motivation Theory," *Journal of Applied Social Psychology*, vol. 30, no. 2, pp. 106-143, 2000.

[19] M. A. Hameed and N. A. G. Arachchilage, "A Model for the Adoption Process of Information System Security Innovations in Organisations: A Theoretical Perspective," In: *The Proceeding of the 27th Australasian Conference on Information Systems (ACIS)*, 2016.

[20] M. A. Hameed and S. Counsell, "Assessing the Influence of Environmental and CEO Characteristics for Adoption of Information Technology in Organizations," *Journal of Technology Management and Innovation*, vol. 7, no. 1, pp. 64-84, 2012.

[21] M. A. Hameed and S. Counsell, "Establishing Relationship between Innovation Characteristics and IT Innovation Adoption in Organisations: A Meta-analysis Approach," *International Journal of Innovation Management*, vol. 18, no. 1, pp. 41, 2014.

[22] M. A. Hameed and S. Counsell, "User Acceptance Determinants of Information Technology Innovation in Organisations," *International Journal of Innovation and Technology Management*, vol. 11, no. 5, pp. 17, 2014.

[23] M. A. Hameed, S. Counsell and S. Swift, "A Meta-analysis of Relationships between Organisational Characteristics and IT Innovation Adoption in Organisations," *Information and Management*, vol. 49, no. 5, pp. 218-232, 2012.

[24] B. Hanus and Y. A. Wu, "Impact of Users' Security Awareness on Desktop Security Behavior: A Protection Motivation Theory Perspective," *Information Systems Management*, vol. 33, No: 1, pp. 2-16, 2016.

[25] T. Herath, R. Chen, J. Wang, K. Banjara, J. Wilbur and H. Rao, "Security Services as Coping Mechanisms: An Investigation into User Intention to Adopt an Email Authentication Service," *Information Systems Journal*, vol. 24, no. 1, pp. 61-84, 2014.

[26] T. Herath and H. R. Rao, "Protection Motivation and Deterrence: A Framework for Security Policy Compliance in Organizations," *European Journal of Information Systems*, vol. 18, no. 2, pp. 106-125, 2009.

[27] J. E. Hunter, F. L. Schmidt and G. B. Jackson, "Meta-Analysis," Beverly Hills, CA: Sage, 1982.

[28] P. Ifinedo, "Understanding Information Systems Security Policy Compliance: An Integration of the Theory of Planned Behaviour and the Protection Motivation Theory," *Computers and Security*, vol. 31, pp. 83-95, 2012.

[29] J. Jansen, and P. Van Schaik, "Persuading End Users to Act Cautiously Online: Initial Findings of a Fear Appeals Study on Phishing," In: *The Proceedings of 11th International Symposium on Human Aspects of Information Security and Assurance (HAISA)*, 2017.

[30] A. C. Johnston, and M. Warkentin, "Fear Appeal and Information Security Behaviors: An Empirical Study," *MIS Quarterly*, vol. 34, no. 3, pp. 549-566, 2010.

[31] F. Lai, D. Li, and C. Hsieh, (2012). "Fighting Identity Theft: The Coping Perspective," Decision Support Systems, vol. 52, pp. 353-363, 2012.

[32] D. Lee, R. Larose, and N. Rifon, "Keeping Our Network Safe: A Model of Online Protection Behaviour," *Behaviour and Information Technology*, vol. 27, no. 5, pp. 445–454, 2008.

[33] Y. Lee, and K. R. Larsen, "Threat or Coping Appraisal: Determinants of SMB Executive's Decision to Adopt Anti-malware Software," *European Journal of Information Systems*, vol. 18, no. 2, pp. 177-187, 2009.

[34] Y. Li, J. Wang, and H. R. Rao, "Adoption of Identity Protection Service: An Integrated Protection Motivation - Precaution Adoption Process Model," In: *The Proceedings of 23rd Americas Conference on Information Systems (AMCIS)*, 2017.

[35] H. Liang, and Y. Xue, "Understanding Security Behaviors in Personal Computer Usage: A Threat Avoidance Perspective," *Journal of the Association for Information Systems*, vol. 11, no. 7, pp. 394-414, 2010.

[36] G. A. Liyanarachchi, "Feasibility of using Student Subjects in Accounting Experiments: A Review," *Pacific Accounting Review*, vol. 19, no. 1, pp. 47-67, 2007.

[37] V. Luu, L. Land and W. Chin, "Safeguarding Against Romance Scams – Using Protection Motivation Theory," In: *The Proceedings of the 25th European Conference on Information Systems (ECIS)*, 2017.

[38] P. Meso, Y. Ding, and S. Xu, "Applying Protection Motivation Theory to Information Security Training for College Students," *Journal of Information Privacy and Security*, vol. 9, no. 1, pp. 47-67, 2013.

[39] N. Mohamed and I. Ahmad, "Information Privacy Concerns, Antecedents and Privacy Measure Use in Social Networking Sites: Evidence from Malaysia," *Computers in Human Behavior*, vol. 28, pp. 2366–2375, 2012.

[40] F. Mwagwabi, T. McGill and M. Dixon, "Short-term and Long-term Effects of Fear Appeals in Improving Compliance with Password Guidelines," *Communications of the Association for Information Systems*, vol. 42, pp 147-182, 2018.

[41] B. Y. Ng, A. Kankanhalli and Y. Xu, "Studying Users' Computer Security Behavior Using the Health Belief Model," *Decision Support Systems*, vol. 46, no. 4, pp. 815-825, 2009.

[42] R.W. Rogers, "A Protection Motivation Theory of Fear Appeals and Attitude Change," *The Journal of Psychology*, vol. 91, pp. 93–114, 1975

[43] R.W. Rogers, "Cognitive and Physiological Processes in Fear Appeals and Attitude Change: A Revised Theory of Protection Motivation," In: J. Cacioppo and R. Petty (Eds.), *Social Psychophysiology. New York: Guilford Press, pp. 153-176*, 1983.

[44] M. L. Sher, P. C. Talley, C. W. Yang, and K. M. Kuo, "Compliance with Electronic Medical Records Privacy Policy: An Empirical Investigation of Hospital Information Technology Staff," *The Journal of Health Care Organization, Provision, and Financing*, vol. 54, pp. 1-12, 2017.

[45] D. Sikolia, D. Twitchell, and G. Sagers, "Protection

Motivation and Deterrence: Evidence from a Fortune 100 Company," *AIS Transactions on Replication Research*, vol. 4, 2018.
[46] M. Siponen, M. A. Mahmood and S. Pahnila, "Employees' Adherence to Information Security Policies: An Exploratory Feld Study," *Information and Management*, vol. 51, pp. 217-224, 2014.
[47] J. Stanton, K. Stam, P. Mastrangelo and J. Jolton, "Analysis of End User Security Behaviors," *Computers and Security*, vol. 24, no, 2, pp 124-133, 2005.
[48] D. W. Sumiyana, "Could Affectivity Compete Better than Efficacy in Describing and Explaining Individuals' Coping Behavior: An Empirical Investigation," *Journal of High Technology Management Research*, vol. 29, pp. 57–70, 2018.
[49] M. B. Tannenbaum, J. Hepler, R. S. Zimmerman, L. Saul, S. Jacobs, K. Wilson and D. Albarracín, "Appealing to Fear: A Meta-analysis of Fear Appeal Effectiveness and Theories," *Psychological Bulletin*, vol. 141, no. 6, pp. 1178-1204, 2015.
[50] N. Thompson, T. J. McGill and X. Wang, "Security Begins at Home: Determinants of Home Computer and Mobile Device Security Behavior," *Computer and Security*, vol. 70, pp. 376-391, 2017.
[51] L. G. Tornatsky and M. Fleischer, "The Process of Technological Innovation," Lexington Books, 1990.
[52] H. Y. S. Tsai, M. Jiang, S. Alhabash, R. LaRose, N. J. Rifon, and S. R. Cotten, "Understanding Online Safety Behaviors: A Protection Motivation Theory Perspective," *Computers and Security*, vol. 59, pp. 138-150, 2016.
[53] C. Z. Tu, J. Adkins, and G. Y. Zhao, "Complying with BYOD Security Policies: A Moderation Model," In: *The Proceedings of the Midwest Association for Information System (MWAIS) 25*, 2018.
[54] A. Vance, M. Siponen and S. Pahnila, "Motivating IS Security Compliance: Insights from Habit and Protection Motivation Theory," *Information and Management*, vol. 49, pp. 190-198, 2012.
[55] S. F. Verkijika, "Understanding Smartphone Security Behaviors: An Extension of the Protection Motivation Theory with Anticipated Regret," *Computer and Security*, (Article in Press), 2018.
[56] M. Warkentin, A. C. Johnston, J. Shropshire and W. D. Barnett, "Continuance of Protective Security Behavior: A Longitudinal Study," *Decision Support Systems*, vol. 92, pp. 25–35, 2016.
[57] M. Workman, W. Bommer, and D. Straub, "Security Lapses and the Omission of Information Security Measures: A Threat Control Model and Empirical Test," *Computers in Human Behavior*, vol. 24, pp. 2799-2816, 2008.
[58] C. Yoon, J. W. Hwang and R. Kim, "Exploring Factors that Influence Students' Behaviors in Information Security," *Journal of Information Systems Education*, vol. 23, no. 4, pp. 407-417, 2012.
[59] X. Zhang, S. Liu, X. Chen, L. Wang, B. Gao and Q. Zhu, "Health Information Privacy Concerns, Antecedents, and Information Disclosure Intention in Online Health Communities," *Information and Management*, vol. 55, pp. 482-493, 2018.
[60] R. W. Zmud, "Diffusion of Modern Software Practices: Influence of Centralization and Formalization," *Management Science*, vol. 28, no.12, pp. 1421–1431, 1982.
[61] N.A.G. Arachchilage and M. Cole, "Designing a mobile game for home computer users to protect against "phishing attacks"," Intenatioal Journal for e- Learning Security (IJeLS), Volume 1, Issue 1/2, March/June 2011.

## Authors' Profiles

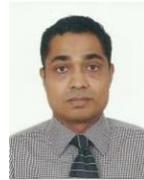

**Mumtaz Abdul Hameed** is currently an Information Technology Consultant and Information Systems Lecturer at the Technovation Consulting and Training Private Limited - Maldives. He received his PhD from Brunel University London - United Kingdom in 2013, after completing his postgraduate studies at the University of Cambridge - United Kingdom. His research interests include: Adoption of IT innovations, Technology acceptance, IS security, Intelligent information processing and IT integration. He has published in Information and Management, Journal of Engineering and Technology Management, among others. Most of his papers were published and presented in the field of innovation adoption in organisations.

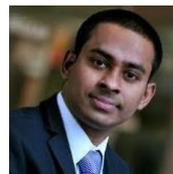

**Nalin Asanka Gamagedara Arachchilage** currently works as a Lecturer in Cyber Security in the Australian Centre for Cyber Security (ACCS) at the University of New South Wales (UNSW Canberra at the Australian Defence Force Academy). He holds a PhD in Usable Security entitled "Security Awareness of Computer Users: A Game Based Learning Approach" from Brunel University London, UK where he developed a game design framework to protect computer users against "phishing attacks". He obtained a BSc (MIS) Hons from University College Dublin, National University of Ireland and has completed a master's degree, MSc in Information Management and Security at the University of Bedfordshire, UK. He is a Sun Certified Java Programmer (SCJP) at Sun Microsystems (now Oracle), USA and professional member of Association for Computing Machinery (ACM).

APPENDIX

| STUDY | YEAR | SAM SUB | SAM SIZE | COUNTRY | THEORY | INN | COR VAL |
|---|---|---|---|---|---|---|---|
| Aurigemma & Mattson [4] | 2014 | IND | 239 | USA | PMT, DT, DTPB | PRC | 0.510 |
| Bélangera et al. [7] | 2017 | SDT | 535 | Singapore | HBM | PRD | 0.400 |
| Bulgurcu et al. [8] | 2010 | ORG | 464 | Malaysia | PMT, SCT, | PRC | 0.419 |
| Burns et al. [9] | 2017 | IND | 377 | USA | PMT | PRC | - |
| Chen [10] | 2017 | SDT | 687 | USA | PMT | PRD | 0.342 |
| Chenoweth et al. [11] | 2009 | IND | 204 | Canada | TPB, RCT | PRC | 0.395 |
| Cho and Ip [12] | 2018 | IND | 418 | South Korea | PMT | PRC | 0.100 |
| Chou and Chien [13] | 2016 | IND | 505 | Canada | TPB, PMT | PRC | 0.320 |
| Crossler [14] | 2010 | IND | 112 | Taiwan | PMT | PRD | 0.050 |
| Hanus and Wu [24] | 2016 | SDT | 229 | USA | PMT | PRD | 0.888 |
| Herath & Rao ([26] | 2009 | ORG | 312 | Finland | TRA | PRC | 0.243 |
| Herath et al. [25] | 2014 | SDT | 134 | USA | PMT | PRD | 0.600 |
| Ifinedo [28] | 2012 | IND | 124 | Finland | PMT | PRC | 0.470 |
| Jansen and Van Schaik ([29] | 2017 | MIX | 1201 | USA | TAM, TTAT | PRD | -0.080 |
| Johnston and Warkentin [30] | 2010 | MIX | 215 | USA | TTAT | PRD | 0.283 |
| Lai et al. [31] | 2012 | SDT | 117 | USA | PMT | PRC | 0.650 |
| Lee et al. [32] | 2008 | SDT | 273 | USA | RCT | PRC | -0.186 |
| Li et al. [34] | 2017 | IND | 616 | USA | PMT | PRD | 0.784 |
| Liang and Xue [35] | 2010 | SDT | 152 | USA | PMT | PRC | 0.260 |
| Luu et al. [37] | 2017 | IND | 399 | USA | PMT | PRD | - |
| Meso et al. [38] | 2013 | SDT | 77 | USA | PMT | PRD | - |
| Mohamed and Ahmad [39] | 2012 | SDT | 340 | USA | PMT | PRC | 0.760 |
| Mwagwabi et al. [40] | 2018 | IND | 194 | USA | PMT, DT, DTPB | PRC | - |
| Ng et al. [41] | 2009 | MIX | 134 | Australia | PMT | PRC | - |
| Sher et al. [44] | 2017 | IND | 310 | USA | PMT | PRC | -0.102 |
| Sikolia et al. [45] | 2018 | IND | 437 | USA | PMT | PRD | 0.000 |
| Siponen et al. [46] | 2014 | IND | 669 | USA | PMT | PRD | 0.140 |
| Sumiyana [48] | 2018 | IND | 580 | South Africa | PMT | PRD | 0.191 |
| Thompson et al. [50] | 2017 | IND | 322 | USA | TPB | PRC | 0.582 |
| Thompson et al. [50] | 2017 | IND | 307 | China | PMT | PRC | 0.157 |
| Tsai et al. [52] | 2016 | IND | 988 | USA | PMT | PRC | 0.070 |
| Tu et al. [53] | 2018 | IND | 122 | Indonesia | PMT | PRD | - |
| Vance et al. [54] | 2012 | IND | 210 | Taiwan | PMT, TRA | PRC | 0.290 |
| Verkijika [55] | 2018 | IND | 385 | Hongkong | TAM, TPB, TTAT | PRC | 0.142 |
| Warkentin et al. [56] | 2016 | SDT | 253 | USA | PMT | PRD | 0.190 |
| Workman et al. [57] | 2008 | IND | 588 | USA | PMT | PRD | - |
| Yoon et al. [58] | 2012 | SDT | 202 | Netherlands | PMT | PRD | - |
| Zhang et al. [59] | 2018 | IND | 337 | USA | DT | PRD | 0.265 |

[Year - Year of study]; [SAM SUB - Sample Subject: ORG - Organisational; IND - Individual; SDT - Student; MIX - Mixed];
[SAM SIZE - Sample Size]; [COUNTRY - Country of study]; [THEORY - Theories used: Protection Motivation Theory - PMT;
Theory of Planned Behaviour - TPB; Technology Threat Avoidance Theory – TTAT; Theory of Reasoned Action - TRA;
Decomposed Theory of Planned Behaviour - DTPB; Deterrence Theory - DT; Technology Acceptance Model - TAM;
Social Cognitive Theory - SCT]; [INN - Type of Innovation: PRC - Process; PRD - Product]; [COR VAL - Correlation Value]